\documentclass[showpacs,preprintnumbers, twocolumn,
amsmath,amssymb,APSl,prd,nofootinbib,superscriptaddress]{revtex4-2} 
\usepackage{graphicx}
\usepackage{hyperref}
\usepackage{tikz}
\usepackage{amsmath,amssymb}
\usepackage{bm}
\usepackage{xcolor}

\definecolor{lime}{HTML}{A6CE39}
\DeclareRobustCommand{\orcidicon}{%
	\begin{tikzpicture}
		\draw[lime, fill=lime] (0,0) 
		circle [radius=0.16] 
		node[white] {{\fontfamily{qag}\selectfont \tiny ID}};
		\draw[white, fill=white] (-0.0625,0.095) 
		circle [radius=0.007];
	\end{tikzpicture}
	\hspace{-2mm}
}

\foreach \x in {A, ..., Z}{%
	\expandafter\xdef\csname orcid\x\endcsname{\noexpand\href{https://orcid.org/\csname orcidauthor\x\endcsname}{\noexpand\orcidicon}}
}

\newcommand\orcidFrancisco{{\href{https://orcid.org/0000-0002-9388-8373}{\orcidicon}}}
\newcommand\orcidManuel{{\href{https://orcid.org/0000-0001-8586-0285}{\orcidicon}}}

\begin{document}
	
\title{Entropy release from Minkowski breaking in regular Schwarzschild black holes}
	
\author{Francisco S. N. Lobo\orcidFrancisco\!\!} 
\email{fslobo@ciencias.ulisboa.pt}
\affiliation{Instituto de Astrof\'{i}sica e Ci\^{e}ncias do Espa\c{c}o, Faculdade de Ci\^{e}ncias da Universidade de Lisboa, Edifício C8, Campo Grande, P-1749-016 Lisbon, Portugal}
\affiliation{Departamento de F\'{i}sica, Faculdade de Ci\^{e}ncias da Universidade de Lisboa, Edif\'{i}cio C8, Campo Grande, P-1749-016 Lisbon, Portugal}
\author{Manuel E. Rodrigues\orcidManuel\!\!} 
\email{esialg@gmail.com}
\affiliation{Faculdade de F\'{i}sica, Programa de P\'{o}s-Gradua\c{c}\~{a}o em F\'{i}sica, Universidade Federal do Par\'{a}, 66075-110, Bel\'{e}m, Par\'{a}, Brazil}
\affiliation{Faculdade de Ci\^{e}ncias Exatas e Tecnologia, Universidade Federal do Par\'{a}, Campus Universit\'{a}rio de Abaetetuba, 68440-000, Abaetetuba, Par\'{a}, Brazil}
	
\date{\today}
	

\begin{abstract}
	The classical formation of a Schwarzschild black hole from a regular, non‑singular configuration has recently been shown to be impossible within general relativity: the geometry inevitably develops a discontinuity at the origin, a phenomenon termed \emph{Minkowski breaking} by Ovalle, Casadio, and Kamenshchik [PRD 113 (2026), 064042]. This obstruction signals that the transition to the Schwarzschild point mass must be a discrete, quantum event. We uncover the thermodynamic footprint of this transition. Using the explicit family of regular Schwarzschild black holes with a de Sitter core, we show that the inner Killing horizon carries a formal Bekenstein‑Hawking entropy $S_{\rm inner} = A_{\rm inner}/4$ that is absent in the singular Schwarzschild state. This entropy is hidden from external observers in equilibrium but, assuming the generalized second law, must be released when the inner horizon disappears. As the collapse parameter $n$ decreases, the inner horizon shrinks and its entropy is gradually released during classical evolution, until the horizon finally vanishes at $n=0$ with the Minkowski breaking. The surface gravity diverges as $n\to0^+$, with the semiclassical description breaking down at $n \sim 1/\ln(h/\ell_P)$; the final disappearance is therefore a deep quantum process. For the $n=3$ regular black hole, the stored entropy is approximately $59\%$ of $A/4$; in the semiclassical limit $n\gg1$, it approaches the full $A/4$. The integer nature of $n$ implies a quantized entropy spectrum, with the Schwarzschild black hole as the ground state within the OCK family. Extensions to $N>1$ families reveal a richer multi-stage cascade, with multiple integer parameters contributing to the entropy release. We discuss how the classical mass‑inflation instability may be circumvented by the quantum disappearance of the Cauchy horizon, and clarify the continuous vs. discrete nature of the collapse. These results suggest a possible realization of horizon area quantization through the discrete internal degrees of freedom of the geometry.
\end{abstract}

\maketitle

\section{Introduction}

The presence of spacetime singularities in classical general relativity is both a foundational challenge and a powerful motivation for the search of a quantum theory of gravity.  The singularity theorems of Penrose, Hawking, and Geroch~\cite{Penrose:1964wq,Hawking:1970zqf,Hawking:1973uf} establish that, under generic conditions, gravitational collapse leads to the formation of curvature singularities where the laws of physics break down.  The Schwarzschild black hole, the simplest and most iconic solution, encapsulates this tension: its interior contains a spacelike singularity hidden behind an event horizon, yet the dynamical process that would produce this exact configuration from a regular initial state has remained elusive.

Over the decades, numerous regular black hole models have been proposed to circumvent the singularity problem, starting with the seminal work of Bardeen~\cite{Bardeen:1968qtr} and followed by many others~\cite{Dymnikova:1992ux,Hayward:2005gi,BenAchour:2020gon}.  These solutions replace the central singularity by a regular core, typically of de Sitter type, and possess an inner (Cauchy) horizon in addition to the outer event horizon.  Despite their mathematical elegance, the physical mechanisms by which such regular configurations could form, and equally important, how they might evolve dynamically, have remained poorly understood.  The Penrose theorem clarifies that singularity avoidance requires the violation of at least one of its assumptions, but it does not prescribe a dynamical path.

A major step forward was recently taken by Ovalle, Casadio, and Kamenshchik (OCK)~\cite{Ovalle:2024wtv,Ovalle:2026lxb}, who constructed an explicit family of regular Schwarzschild interiors.  The geometry is described by a Kerr--Schild metric whose mass function is a polynomial in $r/h$, parametrized by a set of integer exponents $n_i>2$.  The solution matches smoothly to the Schwarzschild exterior at the event horizon, requires no additional charges, and the interior approaches a de Sitter core at the centre.  Remarkably, when the parameters $n_i$ are promoted to functions of time and evolve during collapse, the OCK analysis reveals that the transition from a regular black hole to the singular Schwarzschild point mass cannot be continuous: the metric function develops a discontinuity at the origin when the dominant exponent crosses zero, an effect they termed \emph{Minkowski breaking}.  This kinematic obstruction signals that the formation of a Schwarzschild black hole from a regular state must involve a discrete, quantum event.

The thermodynamic implications of this discovery have not been explored.  The regular black holes of the OCK family possess an inner Killing horizon with a non‑zero surface gravity, and therefore carry a formal geometric entropy $S_{\rm inner} = A_{\rm inner}/4$, in addition to the standard entropy $A/4$ of the event horizon.
In equilibrium, the inner horizon is causally disconnected from the outside world, so its entropy is hidden; it does not contribute to the thermal properties measured by an asymptotic observer. However, when the black hole undergoes a transition that destroys the inner horizon, this hidden entropy, assuming the validity of the generalized second law, must be liberated to the environment to obey the generalised second law of thermodynamics. We therefore interpret $S_{\rm inner}$ as a \emph{potential} gravitational entropy, a bookkeeping device for the internal geometric microstates that become relevant during a non‑adiabatic change.

It is important to stress the logical sequence: the Minkowski breaking is a geometric obstruction that forces the inner horizon to vanish; the release of the hidden entropy then follows as a thermodynamic consequence, assuming the generalized second law. The thermodynamics does not drive the transition; it provides its entropic footprint.
Thus, as the collapse drives the system toward the Schwarzschild ground state within the OCK family, the inner horizon shrinks continuously and its entropy gradually decreases during the classical evolution.  The inner horizon survives down to $n=0$, where it finally vanishes together with the Minkowski breaking: at $n=0$ the metric function at the origin jumps to $f(0)=-1/2 \neq 1$, signalling a loss of local Minkowski structure. As shown below, the surface gravity of the inner horizon diverges as $n\to0^+$, and the semiclassical description breaks down at $n \sim 1/\ln(h/\ell_P)$ for astrophysical black holes, deep in the quantum regime. The entropy $S_{\rm inner}$ is therefore expected to be released through a quantum transition at this point, rather than a classical continuous process.  This suggests a profound connection between the discrete internal structure of regular black holes, the quantization of horizon area, and the origin of black hole entropy.

In this paper, we develop these ideas quantitatively.  Using the simplest member of the OCK family, the $N=1$ solution, we compute the inner horizon radius $h_{\rm c}$ as a function of the integer parameter $n$, and determine the entropy stored in the inner horizon $\Delta S(n) = S_{\rm inner}(n)$.  We show that $\Delta S$ is a positive, finite fraction of the Bekenstein--Hawking entropy; for $n=3$, it amounts to approximately $59\%$ of $A/4$, and in the semiclassical limit $n\gg 1$ the stored fraction approaches the full $A/4$.  The integer nature of $n$ implies that the total entropy of the regular black hole is quantized, forming a discrete spectrum of which the Schwarzschild black hole is the ground state within the OCK family.
We carefully discuss the physical interpretation of the inner‑horizon entropy and temperature, the continuous vs. discrete nature of the collapse parameter, the fulfilment of the second law, and the role of the classical mass‑inflation instability. These results provide a concrete realisation of the idea that black hole entropy may count the number of internal geometric configurations that are indistinguishable from the outside.

The paper is organized as follows. In Sec.~\ref{sec:model} we review the $N=1$ regular Schwarzschild solution and discuss its geometric and matter properties. Section~\ref{sec:inner_entropy} introduces the inner‑horizon entropy and clarifies its thermodynamic status. The continuous classical evolution, the Minkowski breaking, and the associated quantum entropy cascade are analysed in Sec.~\ref{sec:cascade}. The discrete entropy spectrum and its implications for area quantization are presented in Sec.~\ref{sec:quantization}. In the Appendix~\ref{sec:N2}, we extend the analysis to the $N=2$ family, revealing a richer multi-stage cascade where both $n_1$ and $n_2$ transitions contribute to the entropy release, and where the inner horizon radius exhibits algebraic convergence to the $N=1$ limit. We conclude in Sec.~\ref{sec:conclusion} with a summary of our findings and a discussion of future directions.

\section{Regular Schwarzschild black hole with $N=1$}\label{sec:model}

The OCK regular Schwarzschild interior is described by the line element (we set $c=G=1$)
\begin{equation}
	ds^2 = -f(r) dt^2 + \frac{dr^2}{f(r)} + r^2 d\Omega^2,
	\label{metric}
\end{equation}
with
\begin{equation}
	f(r) = \begin{cases}
		1 - \dfrac{2\,m(r)}{r}, & 0 < r \le h,\\[8pt]
		1 - \dfrac{2\mathcal{M}}{r}, & r > h,
	\end{cases}
	\label{f_def}
\end{equation}
where $h=2\mathcal{M}$ is the event horizon radius and $\mathcal{M}$ is the ADM mass.  The Misner–Sharp mass $m(r)$ for the $N=1$ case is given by~\cite{Ovalle:2026lxb}
\begin{equation}
	m(r) = \frac{r}{2(n-2)}\left[\frac{r^2}{h^2}\,(n+1) - 3\left(\frac{r}{h}\right)^n\right],\qquad n>2,
	\label{m_N1}
\end{equation}
which satisfies the matching conditions $m(h)=h/2=\mathcal{M}$ and $m'(h)=0$.  The corresponding metric function is
\begin{equation}
	f(r) = 1 - \frac{1}{n-2}\left[\frac{r^2}{h^2}\,(n+1) - 3\left(\frac{r}{h}\right)^n\right].
	\label{f_N1}
\end{equation}
At the origin, $f(0)=1$ and the geometry approaches a de Sitter core with an effective cosmological constant
\begin{equation}
	\Lambda_{\rm eff} = \frac{3}{h^2}\,\frac{n+1}{n-2}.
	\label{Lambda}
\end{equation}

The Misner–Sharp mass $m(r)$ represents the total energy enclosed within a sphere of radius $r$ and, for $n>2$, it is a monotonically increasing function of $r$.  Its leading behaviour at small $r$,
\begin{equation}
	m(r \to 0) \simeq \frac{n+1}{2(n-2)}\,\frac{r^3}{h^2},
\end{equation}
confirms the de Sitter core: the energy density $\rho = m'(r)/(4\pi r^2)$ tends to the constant value $\rho_0 = 3(n+1)/[8\pi h^2(n-2)]$, reproducing the interior cosmological constant \eqref{Lambda}.  This is the characteristic property that renders the geometry regular at the centre; the Kretschmann scalar indeed remains finite as $r\to 0$,
\begin{equation}
	K(r=0) = \frac{24\,(n+1)^2}{h^4\,(n-2)^2}.
\end{equation}

For $n>2$, the metric function $f(r)$ possesses a second zero at $r = h_{\rm c} < h$, which corresponds to an inner (Cauchy) horizon.  An analytic closed form for $h_{\rm c}$ is not available for generic $n$, but special cases and asymptotic limits are instructive.  For $n=3$, the equation $f(r)=0$ reduces to the cubic $3(r/h)^3 - 4(r/h)^2 + 1 = 0$, whose root inside the event horizon is
\begin{equation}
	h_{\rm c}(n=3) = \frac{1+\sqrt{13}}{6}\,h \approx 0.7676\,h,
	\label{n3_horizon}
\end{equation}
while for $n\gg 1$ the metric approaches the pure de Sitter form $f(r) \simeq 1 - r^2/h^2$, so that $h_{\rm c} \simeq h$.
As $n$ decreases, the inner horizon shrinks.  At $n=2$ the geometry develops a logarithmic curvature singularity, but the inner horizon persists.  The inner horizon finally vanishes at $n=0$, precisely where the Minkowski breaking takes place: the metric function at the origin jumps to $f(0) = -1/2 \neq 1$, signalling a loss of local Minkowski structure.

The matter content that supports this geometry can be read off from the Einstein equations.  Expressed as an anisotropic fluid, the energy-momentum tensor has the form $T^\mu_\nu = {\rm diag}(-\rho, p_r, p_t, p_t)$ with the radial pressure equal to the negative of the energy density, $p_r = -\rho$, and the transverse pressure given by the conservation equation $p_t = -\rho - \frac{r}{2}\rho'$.  Using the energy density
\begin{equation}
	\rho(r) = \frac{3}{8\pi h^2}\frac{n+1}{n-2}\left[1 - \left(\frac{r}{h}\right)^{n-2}\right],
	\label{rho}
\end{equation}
which is everywhere positive for $n>2$, we obtain
\begin{equation}
	\label{pt_correct}
	p_t(r) = \frac{3(n+1)}{16\pi (n-2) h^2}
	\left[ -2 + n \left(\frac{r}{h}\right)^{\!n-2} \right].
\end{equation}
The fluid automatically obeys the null convergence condition because $\rho + p_r = 0$ and
\begin{equation}
	\rho + p_t = \frac{3(n+1)}{16\pi h^2}\left(\frac{r}{h}\right)^{n-2} \ge 0,
\end{equation}
for all $r\le h$.

The strong energy condition (SEC), which requires $\rho + p_r + 2p_t \ge 0$, reduces to $p_t \ge 0$ since $p_r = -\rho$.  From Eq.~\eqref{pt_correct}, the SEC is violated for $(r/h)^{n-2} < 2/n$, i.e. sufficiently close to the centre.  For the illustrative case $n=3$, the violation occurs for $r/h < 2/3 \approx 0.667$.  This behaviour is typical of regular black holes with a de Sitter core, where the negative tangential pressure supports the core against gravitational collapse.  For larger $n$, the violation region shrinks; in the semiclassical limit $n\gg1$, the SEC is violated only in a tiny neighbourhood of the origin, consistent with the approach to a pure de Sitter spacetime.

The parameter $n$ is an integer ($n>2$) for the static regular solutions.  In the dynamical setting, one promotes the geometry to a time‑dependent one by introducing the Eddington–Finkelstein coordinate $v$ through $dv = dt + dr/f$, so that $f = f(v,r)$ and $n = n(v)$.  The function $n(v)$ evolves according to the collapse dynamics; the null convergence condition $R_{\mu\nu}\ell^\mu\ell^\nu \ge 0$ (with $\ell^\mu$ a null vector) imposes $\dot{n}(v) < 0$, i.e., $n$ must decrease during collapse~\cite{Ovalle:2026lxb}.  This evolution drives the system from an initial regular state ($n>2$) toward increasingly singular configurations.  The Minkowski breaking occurs precisely at $n=0$: the metric function at the origin becomes $f(0) = -1/2 \neq 1$, signalling a discontinuous change in the local geometry.  The spacetime ceases to be locally Minkowskian at $r=0$, and the transition from a regular interior to the Schwarzschild point mass cannot be accomplished continuously.

\section{Inner‑horizon entropy: thermodynamic interpretation}\label{sec:inner_entropy}

For $n>2$, the metric function $f(r)$ possesses a second root at $r=h_{\rm c}<h$, corresponding to an inner (Cauchy) horizon.  Since $f(r)$ vanishes at $h_{\rm c}$ and the surface gravity
\begin{equation}
	\kappa_{\rm inner} = \frac12\,f'(h_{\rm c})
\end{equation}
is non‑zero (as can be checked numerically for any finite $n>2$), this is a Killing horizon and one may formally associate with it a Bekenstein--Hawking entropy
\begin{equation}
	S_{\rm inner} = \frac{A_{\rm inner}}{4} = \pi h_{\rm c}^2.
	\label{S_inner}
\end{equation}

In Einstein gravity, the Wald--Noether‑charge formalism reduces to the area law $S = A/4$ for any Killing horizon, including the inner one~\cite{Wald:1993nt,Iyer:1994ys,Wald:1999vt}.  We adopt this identification as a working assumption, while bearing in mind that the physical interpretation of the inner horizon entropy differs fundamentally from that of the event horizon.

In standard black‑hole thermodynamics one considers only the region exterior to the event horizon and assigns the entropy $S = A/4$ to the black hole as seen by an external observer.  The inner (Cauchy) horizon is not in causal contact with the outside world, so its area does not contribute to the entropy measured at infinity, as long as the black hole remains in equilibrium.

Our proposal is different: we are considering a \emph{dynamical process} in which a regular black hole (which possesses an inner horizon) evolves into a Schwarzschild black hole (which does not).  During this process the inner horizon shrinks and finally disappears.  The geometric entropy $\pi h_{\rm c}^{2}$ that was stored in the inner horizon, under the assumption that the generalized second law remains valid, cannot simply vanish; it must be \emph{released} to the environment (as Hawking radiation, gravitational waves, or some other form of energy) in order not to violate the generalised second law.  In this sense, the inner‑horizon entropy is a \emph{hidden} contribution to the total gravitational entropy of the system, which becomes observable only when the black hole undergoes a transition to the Schwarzschild state.  The additive rule
\begin{equation}
	S_{\rm reg} = S_{\rm outer} + S_{\rm inner} = \frac{A}{4} + \pi h_{\rm c}^{2}
	\label{Sreg}
\end{equation}
is therefore a natural bookkeeping device for the total number of geometric microstates that are consistent with the exterior Schwarzschild geometry.

From an external perspective, only the event‑horizon entropy is accessible as a coarse‑grained thermodynamic entropy; the inner‑horizon entropy becomes relevant only when the horizon is destroyed, effectively resolving additional microscopic degrees of freedom.

The surface gravity $\kappa_{\rm inner}$ defines a formal temperature $T_{\rm inner} = \hbar\kappa_{\rm inner}/(2\pi)$.  The region between the two horizons has a ``reversed'' signature, and the Cauchy horizon, although a Killing horizon, does not radiate towards a static observer in the ordinary sense.  Consequently, $T_{\rm inner}$ is a purely geometric quantity, and the entropy $S_{\rm inner}$ is a geometric measure of the horizon area.  Our use of these quantities is therefore at the level of a \emph{classical geometric counting of states}, not a statement about a thermal bath that a local observer would detect.  The physical importance of $S_{\rm inner}$ emerges only when the inner horizon is destroyed and its entropy is converted into radiation that can reach infinity.  The divergence of $T_{\rm inner}$ as $n\to 0^{+}$ signals that the semiclassical description breaks down near the Minkowski breaking point, which is consistent with the idea that the final stage is a quantum event.

One may worry that the total entropy of a spacetime with multiple horizons is not simply the sum of the individual Bekenstein-Hawking terms.  In the absence of a complete quantum gravity theory, the additive rule $S_{\rm reg} = S_{\rm outer} + S_{\rm inner}$ is the simplest and most natural choice.  It corresponds to the assumption that the microstates of the two horizons are independent, so that the total number of microstates factorises, $\Omega_{\rm total} = \Omega_{\rm outer}\,\Omega_{\rm inner}$, leading to $S = \ln\Omega_{\rm total} = S_{\rm outer} + S_{\rm inner}$.  Alternative rules (e.g. subtractive) would imply strong correlations between the two horizons for which there is no evidence in the present kinematic model. We therefore adopt the additive rule as a working hypothesis throughout this work.

\section{Minkowski breaking and the quantum entropy cascade}\label{sec:cascade}

\subsection{The second law}

The total entropy of the regular black hole is $S_{\rm reg} = A/4 + \pi h_{\rm c}^{2}$, while the Schwarzschild black hole has $S_{\rm Sch} = A/4$.  The difference
\begin{equation}
	\Delta S(n) = S_{\rm reg}(n) - S_{\rm Sch} = \pi h_{\rm c}^{2}
	\label{DeltaS}
\end{equation}
is the entropy that, if the generalized second law is to hold, must be liberated when the inner horizon disappears.

To determine $h_{\rm c}$ as a function of $n$, one solves $f(r)=0$ for $r<h$.  Although a closed analytic expression is not available for generic $n$, several limits are informative.  For $n\gg1$ the metric approaches the pure de Sitter core with $f(r)\simeq 1 - r^2/h^2$, so that $h_{\rm c}\simeq h$ and $\Delta S \to A/4$: highly excited regular black holes have an inner horizon nearly coincident with the event horizon, and their total entropy is nearly twice the Schwarzschild value.  For the illustrative case $n=3$, the equation $f(r)=0$ reduces to $3(r/h)^3 - 4(r/h)^2 + 1 = 0$.  Factorising out the event horizon root $(r/h=1)$ gives the quadratic $3(r/h)^2 - r/h - 1 = 0$, whose positive solution is given by Eq.~(\ref{n3_horizon}).
Hence
\begin{equation}
	\Delta S(n=3) = \pi h_{\rm c}^2 \approx 0.589\,\pi h^2 = 0.589\,\frac{A}{4}.
	\label{deltaS_n3}
\end{equation}
Thus, a regular black hole with $n=3$ stores an additional entropy equal to approximately $59\%$ of the Bekenstein--Hawking entropy of the event horizon.

As $n$ decreases from a regular value toward zero, the inner horizon shrinks continuously and $\Delta S$ decreases correspondingly.  The inner horizon survives down to $n=0$, where it finally vanishes together with the Minkowski breaking: at $n=0$ the metric function at the origin jumps to $f(0)=-0.5\neq1$, signalling a classical discontinuity in the geometry.  The spacetime ceases to be locally Minkowskian at $r=0$, and the entropy $S_{\rm inner}$ is expected to be released abruptly at this point.  This classical obstruction indicates that the entire collapse process, including the cascade through the discrete $n$ levels that releases the entropy $\Delta S(n)$, must be fundamentally quantum in nature.

The parameter $n(v)$ in the dynamical collapse is a continuous function of the advanced time $v$; it is not restricted to integer values.  The integer values label the \emph{static} regular solutions, which form a discrete family of equilibrium states.  During a classical collapse, $n(v)$ passes through all real values, so the evolution is continuous in that sense.  The quantisation we refer to applies to the equilibrium states, not to the instantaneous value of $n(v)$.  One may picture the dynamical cascade as a sequence of quantum jumps between adjacent integer levels, but the classical continuous evolution does not pause at integer values.  Thus there is no contradiction between a discrete entropy spectrum and a continuous time dependence of $n(v)$ before the critical point.

If one considers only the black hole itself, then $\Delta S = S_{\rm Sch} - S_{\rm reg} = -\pi h_{\rm c}^{2} < 0$, which would appear to violate the second law.  However, the collapse is not an adiabatic process.  The regular black hole is an excited state; as it decays to the ground state (Schwarzschild), it \emph{emits} the excess entropy into the environment.  The total entropy of the universe increases because the emitted radiation carries an entropy $\Delta S_{\rm rad} \ge \pi h_{\rm c}^{2}$.  The black hole entropy decreases, but the entropy of the surroundings increases by at least the same amount, so the generalised second law is satisfied.

This is completely analogous to a hot body cooling by radiation: the body's entropy decreases, but the total entropy (body + radiation) increases.  In the Euclidean path‑integral picture, the transition probability is enhanced by the factor $\exp[\Delta S]$ precisely because the process is thermodynamically favoured once the emitted entropy is accounted for. This entropy release is an irreversible process, reflecting the thermodynamic arrow of time: the transition from a regular excited state to the Schwarzschild ground state increases the total entropy of the Universe, consistent with the second law of thermodynamics. The irreversibility is encoded in the exponential enhancement of the transition probability, which favours decay toward the lower-entropy ground state while the emitted radiation carries away the excess entropy.

\subsection{Mass inflation and the quantum disappearance of the Cauchy horizon}

A well‑known obstacle for regular black holes is the mass‑inflation instability: the inner (Cauchy) horizon is classically unstable, and small perturbations cause a diverging exponential growth of the mass function, likely turning the Cauchy horizon into a null singularity.  In the OCK scenario, the inner horizon shrinks and vanishes \emph{before} the full nonlinear development of the mass inflation can occur.  The divergence of the inner‑horizon temperature as $n\to 0^{+}$ indicates that quantum effects become dominant at that stage.  It is therefore plausible that the disappearance of the Cauchy horizon is not a purely classical process but rather a quantum transition, which would naturally circumvent the mass‑inflation instability.  A complete treatment would require a quantum gravitational or semiclassical analysis, which lies beyond the scope of the present work.

In a full quantum treatment, the evolution from a regular state to the Schwarzschild ground state could correspond to a sequence of tunneling events between discrete geometric microstates. In the standard semiclassical framework, one expects the probability of a transition from a state with parameter $n$ to the ground state to be related to the difference of Euclidean actions $\Delta I_{\rm E}$ by $\Gamma \propto \exp[-\Delta I_{\rm E}]$. If the Euclidean action takes the familiar form $I_{\rm E} = \beta \mathcal{M} - S$, then $\Gamma \propto \exp[\Delta S]$~\cite{Hawking:1975vcx}. We stress that no instanton describing the transition has been constructed; the relation $\Gamma \propto e^{\Delta S}$ is an expectation based on the standard semiclassical framework, not an established result.

With $\Delta S>0$, the transition is exponentially favoured, in full agreement with the second law of thermodynamics.  Physically, the entropy release could manifest itself as a burst of Hawking radiation or gravitational waves, carrying away the energy and information that were stored in the inner horizon. A detailed modelling of the emission process is beyond the scope of this work; here we only note that the entropy is expected to be released through a non‑equilibrium process. If the transition between the discrete static equilibrium states (integer $n$) occurs via quantum jumps, the entropy would be released in discrete steps. In a purely classical continuous evolution, however, $n(v)$ passes through all real values and the entropy release would be continuous. The discrete spectrum therefore applies to the equilibrium states, while the dynamical trajectory may be continuous.

Finally, we note that the inner horizon surface gravity (taking the positive magnitude to define a positive temperature) is
\begin{eqnarray}
	\kappa_{\rm inner} &=& -\frac12 f'(h_{\rm c}) 
	\nonumber \\
	&=& \frac{1}{h(n-2)}\left[\frac{3n}{2}\left(\frac{h_{\rm c}}{h}\right)^{n-1} - (n+1)\frac{h_{\rm c}}{h}\right],
\end{eqnarray}
which provides an effective temperature $T_{\rm inner} = \hbar\kappa_{\rm inner}/(2\pi)$. The sign convention is chosen so that $\kappa_{\rm inner}$ is positive; the signed derivative $f'(h_c)$ is negative for the Cauchy horizon.

As $n\to 0^{+}$, the inner horizon radius vanishes as $h_c/h \simeq (2/3)^{1/n}$, and the surface gravity diverges as $\kappa_{\rm inner} \sim \frac{n}{2h}(3/2)^{1/n}$. The semiclassical description breaks down when this exceeds the Planck temperature $T_{\rm P} = \sqrt{\hbar c^5/G}$. Solving $\kappa_{\rm inner} \sim 1/\ell_{\rm P}$ yields an order-of-magnitude estimate $n \sim 1/\ln(h/\ell_{\rm P})$. For astrophysical black holes, $h/\ell_{\rm P} \sim 10^{38}$, so this occurs at $n \sim 0.01$, deep in the quantum regime.

\section{Discrete entropy spectrum and area quantization}\label{sec:quantization}

The integer character of the parameter $n$, inherited from the construction of the OCK solution as a finite polynomial in $r/h$, has a direct thermodynamic consequence: the inner horizon radius $h_{\rm c}(n)$ and the associated entropy $S_{\rm inner}(n)$ can only take a discrete set of values.  The regular black holes therefore form a discrete family of states, each characterised by its total entropy $S_{\rm reg}(n) = A/4 + S_{\rm inner}(n)$.  Table~\ref{tab:entropy} lists the inner horizon radius, the entropy fraction $\Delta S / (A/4)$, and the associated surface gravity and temperature for the first few values of $n$.

\begin{table}[h]
	\caption{Inner horizon radius, entropy release, surface gravity (positive magnitude), and effective temperature for the $N=1$ regular black holes.  $h_{\rm c}$ is obtained by numerically solving $f(h_{\rm c})=0$ with Eq.~\eqref{f_N1}.  $\kappa_{\rm inner}$ and $T_{\rm inner}$ are given in units of $1/h$ and $\hbar/h$, respectively.}
	\label{tab:entropy}
	\begin{ruledtabular}
		\begin{tabular}{c c c c c c}
			$n$ & $h_{\rm c}/h$ & $\Delta S / (A/4)$ & $S_{\rm reg}/(A/4)$ & $\kappa_{\rm inner} h$ & $T_{\rm inner} h/\hbar$ \\
			\hline
			$3$   & $0.7676$ & $0.589$ & $1.589$ & $0.419$ & $0.0667$ \\
			$4$   & $0.8165$ & $0.667$ & $1.667$ & $0.408$ & $0.0650$ \\
			$5$   & $0.8484$ & $0.720$ & $1.720$ & $0.402$ & $0.0640$ \\
			$6$   & $0.871$  & $0.759$ & $1.759$ & $0.397$ & $0.0632$ \\
			$7$   & $0.8875$ & $0.788$ & $1.788$ & $0.393$ & $0.0625$ \\
			$10$  & $0.919$  & $0.845$ & $1.845$ & $0.384$ & $0.0611$ \\
			$\infty$ & $1$ & $1$ & $2$ & $0$ & $0$ \\
		\end{tabular}
	\end{ruledtabular}
\end{table}

\begin{figure}
	\centering 
	\includegraphics[width=0.475\textwidth]{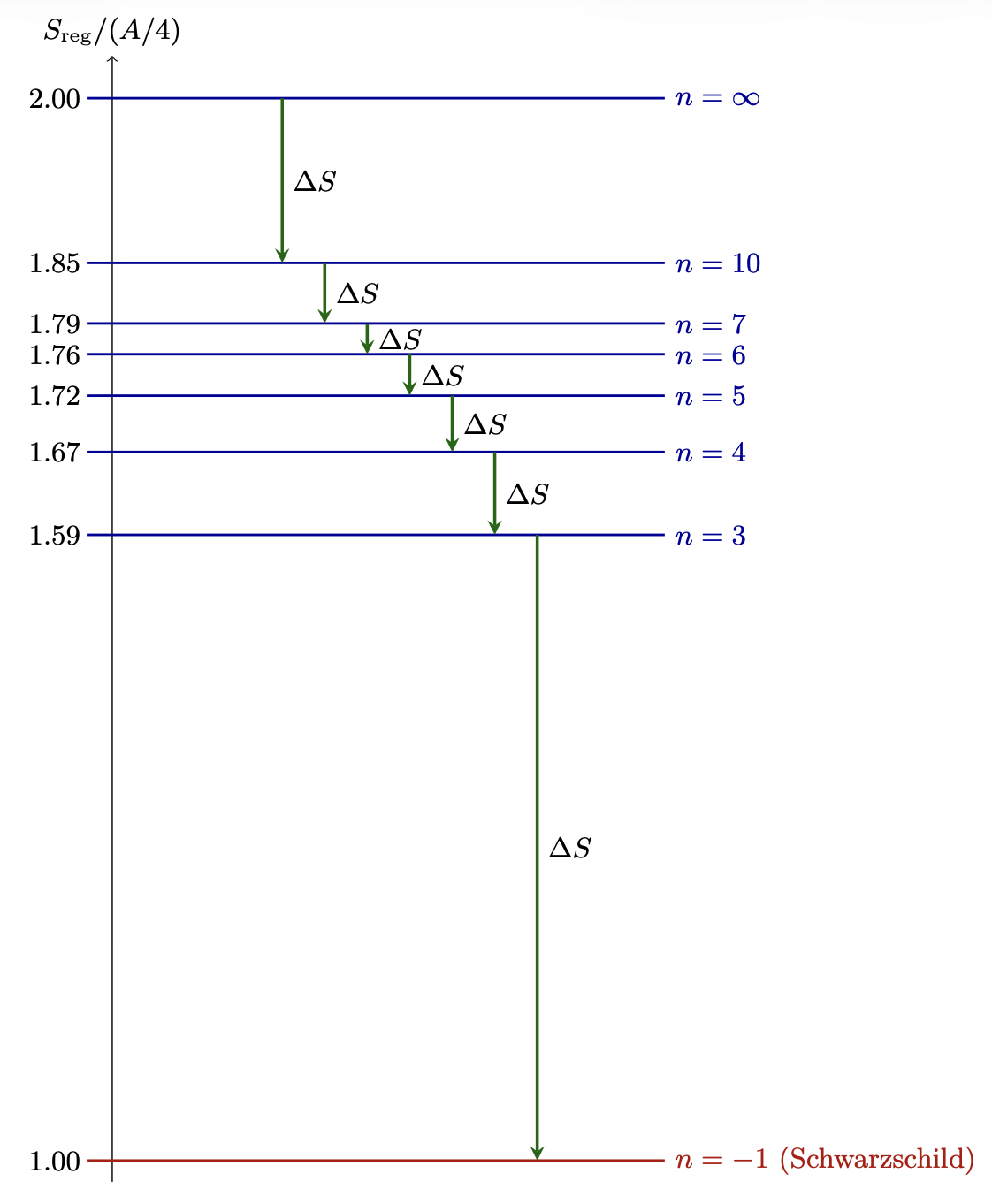}
	\caption{Discrete total entropy $S_{\rm reg}(n)=\pi h^2 + \pi h_{\rm c}^2(n)$ for the $N=1$ regular black hole.  The ground state within the OCK family ($n=-1$, Schwarzschild) has $S=A/4$.  Arrows indicate allowed transitions during collapse, releasing entropy $\Delta S$.}
	\label{fig:entropy}
\end{figure}

Figure~\ref{fig:entropy} gives a schematic representation of these energy levels.  As $n$ decreases, the inner horizon shrinks and the total entropy $S_{\rm reg}$ drops by finite amounts, until it reaches the ground state within the OCK family ($n=-1$, Schwarzschild) with entropy $S = A/4$.

\subsection{Physical interpretation: regular black holes as excited states}

The discrete entropy spectrum suggests that the regular black holes are excited geometric configurations of the Schwarzschild black hole. Each state has the same mass and the same exterior geometry, yet it stores additional information in the internal structure, which is encoded in the parameter $n$ and reflected in the extra entropy $\Delta S(n)$.  The ground state within the OCK family ($n=-1$) is the Schwarzschild black hole, where all the mass is concentrated at the singularity and no inner horizon exists.

From this perspective, the dynamical evolution $\dot n < 0$ driven by the null convergence condition corresponds to a cascade down the entropy ladder: the system decays from states of higher $n$ (larger inner horizon, larger $\Delta S$) to states of lower $n$ (smaller inner horizon, smaller $\Delta S$). If the transition between the discrete static equilibrium states (integer $n$) occurs via quantum jumps, each step releases an amount of entropy $\delta S = S_{\rm reg}(n) - S_{\rm reg}(n-1)$. In a purely classical continuous evolution, however, $n(v)$ passes through all real values and the entropy release is continuous; the discreteness applies to the spectrum of equilibrium states, not necessarily to the dynamical trajectory. The Minkowski breaking at $n=0$ is the final step of this cascade, where the inner horizon finally vanishes and the geometry becomes indistinguishable from the Schwarzschild exterior.  The classical discontinuity at this point signals the impossibility of a continuous transition to the Schwarzschild point mass.

\subsection{Connection to Bekenstein's area quantization}

The discreteness of the parameter $n$ is reminiscent of Bekenstein's old proposal that the horizon area is quantized~\cite{Bekenstein:1974jk}.  In the present model, the event horizon area $A=4\pi h^2$ is fixed because the total mass $\mathcal{M}$ is held constant.  However, the inner horizon area $A_{\rm inner}=4\pi h_{\rm c}^2$ takes a discrete set of values, and one may view the total entropy
\begin{equation}
	S_{\rm reg} = \frac{A}{4} + \frac{A_{\rm inner}}{4} = \frac{A}{4}\bigl[1 + \mathcal{F}(n)\bigr],
	\qquad \mathcal{F}(n) \equiv \frac{h_{\rm c}^2(n)}{h^2}
\end{equation}
as a quantized quantity.  In the semiclassical limit $n\gg1$, $h_{\rm c}\to h$ and $\mathcal{F}(n)\to1$, so that $S_{\rm reg}\to 2(A/4)$.  This suggests that highly excited regular black holes have \emph{twice} the entropy of the Schwarzschild black hole, and that the transition from the most excited state to the ground state liberates an entropy equal to the entire Bekenstein--Hawking value $A/4$.

It is important to stress that the discrete spectrum obtained here does not reproduce Bekenstein's equally spaced area spectrum; rather, it provides a different kind of quantization that stems from the discrete family of interior solutions. Nevertheless, this observation provides a concrete suggestion that the Bekenstein--Hawking entropy may count the number of internal geometric configurations that are indistinguishable from the outside. In the OCK family, the integer $n$ labels precisely such configurations: each value of $n$ gives a different interior, but all of them share the same exterior Schwarzschild metric.

\subsection{Thermodynamic arrow and transition probability}

Because \(\Delta S(n) > 0\) for all regular states (\(n > 0\)) where the inner horizon exists, a transition from a regular state to the Schwarzschild state (or to any state with a smaller \(n\)) increases the entropy of the environment and is therefore thermodynamically favoured. In a semiclassical quantum description, one expects the probability of tunneling from a state with parameter \(n\) to the ground state to be related to the difference of Euclidean actions by~\cite{Hawking:1975vcx}:
\begin{equation}
	\Gamma_{n\to -1} \propto \exp\bigl[-\Delta I_{\rm E}\bigr] .
\end{equation}
If the Euclidean action takes the standard form \(I_{\rm E} = \beta \mathcal{M} - S\), then \(\Gamma_{n\to -1} \propto \exp[\Delta S(n)]\).
Using the values in Table~\ref{tab:entropy}, this probability is exponentially large for all \(n\ge 3\), e.g., \(\Gamma_{3\to -1} \propto e^{0.589 A/4}\), and becomes even larger for higher \(n\).  Thus, once the collapse has proceeded to values of \(n\) of order a few, the system is overwhelmingly likely to tunnel to the Schwarzschild ground state.

This tunneling process is the quantum analogue of the classical Minkowski breaking.  Instead of a discontinuous jump in the metric, quantum mechanics allows a smooth but exponentially fast transition, during which the entropy \(S_{\rm inner}\) is carried away by emitted radiation.  The energy required for this radiation must ultimately come from the mass of the black hole, which in a realistic collapse is not strictly fixed.  A fully self‑consistent treatment would therefore require coupling the kinematic model to the semiclassical Einstein equations, allowing the mass to decrease as Hawking radiation is emitted.

\subsection{Limitations and the need for a finite spectrum}

A note of caution is in order.  The integer \(n\) can be arbitrarily large in the classical construction, leading to an infinite tower of regular black hole states with entropy approaching \(2(A/4)\) from above.  An infinite number of distinct microstates for a fixed macroscopic mass would imply an unbounded density of states, which is likely unphysical in a quantum theory of gravity.  In practice, one expects that quantum gravitational effects will impose a maximum allowed value of \(n\), perhaps through a Planck‑scale cutoff on the effective cosmological constant or on the curvature at the centre.  The existence of such a cutoff would render the entropy spectrum finite, which appears necessary if the number of internal states is to remain finite for a fixed ADM mass, with the most excited state having an entropy close to, but strictly less than, \(2(A/4)\). This strongly suggests that the classical OCK family cannot be the complete quantum description and that an ultraviolet cutoff is required for a consistent statistical interpretation.  The exact value of the cutoff is a question for a future quantum treatment.

Despite this caveat, the present analysis demonstrates that the kinematic discreteness inherent in the OCK solution naturally leads to a quantized entropy spectrum, offering a novel and compelling perspective on the origin of black hole entropy and on the role of internal degrees of freedom in gravitational collapse.

\subsection{Extension to the $N>1$ family}

The $N=1$ solution discussed so far is the simplest member of a much richer family of regular Schwarzschild black holes constructed in~\cite{Ovalle:2026lxb}. For general $N$, the interior Misner--Sharp mass function is given by
\begin{eqnarray}
	m(r) &=& \frac{r}{2}\Bigg[ \left(\frac{r}{h}\right)^{\!2} \prod_{i=1}^{N}\frac{n_i+1}{n_i-2}
	+ 3(-1)^N  \times
	\nonumber \\
	&& \sum_{k=1}^{N} \frac{1}{n_k-2}\left(\frac{r}{h}\right)^{\!n_k}
	\prod_{\substack{i=1\\ i\neq k}}^{N} \frac{n_i+1}{n_k-n_i} \Bigg],
	\label{m_N}
\end{eqnarray}
where the exponents are ordered integers $2 < n_1 < n_2 < \dots < n_N$.  For $N=1$, the product term in the sum reduces to unity (empty product) and Eq.~\eqref{m_N} recovers the $N=1$ mass function~\eqref{m_N1}.  The metric function retains the form~\eqref{f_N1} with $m(r)$ given by~\eqref{m_N}.

The inner horizon radius $h_{\rm c}=h_{\rm c}(\{n_i\})$ is the smallest positive root of $f(r)=0$ besides the event horizon $r=h$, and must be determined numerically for generic $\{n_i\}$.  The total entropy of the regular black hole is then
\begin{equation}
	S_{\rm reg}(\{n_i\}) = \frac{A}{4} + \pi h_{\rm c}^{\,2}(\{n_i\}),
	\label{SregN}
\end{equation}
which depends on the full set of $N$ integers.  Consequently, the entropy spectrum is far more intricate than in the $N=1$ case: each ordered tuple $(n_1,\dots,n_N)$ that yields a regular interior (i.e., $f(r)>0$ for $0<r<h$ and a well-defined inner horizon) defines a distinct geometric microstate.  The number of such microstates for a fixed ADM mass grows rapidly with $N$, providing a large degeneracy of configurations that share the same exterior Schwarzschild metric.

During a dynamical collapse, the exponents become functions of the advanced time $v$, $n_i=n_i(v)$, and the null convergence condition enforces $\dot n_i(v) < 0$, so that all $n_i$ decrease monotonically.  The ordering $n_1(v_0)<n_2(v_0)<\dots<n_N(v_0)$ is preserved by the constraint $\dot n_i(v)\le\dot n_j(v)$ for $i<j$~\cite{Ovalle:2026lxb}.  The smallest exponent $n_1(v)$ is the first to reach the critical values: at $n_1=2$ a logarithmic curvature singularity appears, yet the inner horizon survives; at $n_1=0$ the metric function develops the Minkowski breaking discontinuity at the origin, and the inner horizon vanishes entirely, exactly as in the $N=1$ case. The remaining exponents $n_2,\dots,n_N$ continue to decrease after this point, but the geometry has already become singular, and the thermodynamic description based on the inner horizon is no longer applicable.

Nevertheless, the entropy release during the regular phase is significantly richer: if the transition between the discrete static equilibrium states (integer values of \(n_1\)) occurs via quantum jumps, then as \(n_1(v)\) passes through each integer value, the inner horizon area changes by a finite amount and a quantum of entropy \(\delta S = S_{\rm reg}(n_1) - S_{\rm reg}(n_1-1)\) (with the other \(n_i\) fixed at their initial large values) is emitted. In a purely classical continuous evolution, however, \(n_1(v)\) passes through all real values and the entropy release is continuous; the discreteness applies to the spectrum of equilibrium states, not necessarily to the dynamical trajectory. This multi-stage cascade is a distinctive prediction of the $N>1$ models and could lead to a more complex signature than the single $N=1$ ladder.

In the semiclassical limit $N\gg 1$ and $n_i\gg 1$, the inner horizon approaches the event horizon, $h_{\rm c}\to h$, and the total entropy saturates at $S_{\rm reg}\to 2(A/4)$.  The transition from the most excited state (all $n_i\gg 1$) to the Schwarzschild ground state then liberates an entropy equal to the entire Bekenstein--Hawking value $A/4$.  Furthermore, the integer labels $\{n_i\}$ provide a natural discretisation of the interior geometry: the number of distinct regular states $\Omega({\cal M})$ for a fixed macroscopic mass can be computed combinatorially from the allowed ordered tuples, and in the semiclassical limit one expects $S_{\rm BH} = \ln\Omega({\cal M})$.  This offers a concrete statistical-mechanics proposal, conjecturing that black hole entropy could be understood as the logarithm of the number of internal geometric configurations that are indistinguishable from the outside.

\section{Conclusions}\label{sec:conclusion}

We have demonstrated that the regular Schwarzschild black holes constructed by Ovalle, Casadio, and Kamenshchik carry an additional formal Bekenstein--Hawking entropy associated with their inner Killing horizon, \(S_{\rm inner} = \pi h_{\rm c}^2\). While this entropy is hidden from an asymptotic observer in equilibrium, it, under the assumption of the generalized second law, must be released when the inner horizon is destroyed.  We have clarified the thermodynamic role of the inner horizon, the continuous nature of the collapse parameter, and the way the classical mass‑inflation instability might be bypassed by the quantum nature of the final transition.
Using the explicit \(N=1\) solution, we computed the entropy stored in the inner horizon for the \(n=3\) regular black hole to be \(\Delta S \approx 0.59\,(A/4)\); in the limit \(n\gg 1\), where the inner horizon approaches the event horizon, the stored fraction approaches the full Bekenstein--Hawking value \(A/4\).  The integer nature of the parameter \(n\) implies that the total entropy \(S_{\rm reg}(n) = A/4 + \pi h_{\rm c}^2(n)\) is quantized, suggesting a connection between the discrete internal structure of regular black holes and the old idea of horizon area quantization.

Our results support the interpretation that the formation of a Schwarzschild black hole from a regular state is not a classical continuous process but rather a quantum transition between discrete geometric microstates.  The regular black holes can be viewed as excited geometric configurations of the Schwarzschild black hole, with the integer \(n\) playing the role of a quantum number that labels the internal geometry. During a classical collapse, the parameter \(n(v)\) evolves continuously through all real values, so the entropy release is continuous in time. If, however, the transition between the discrete static equilibrium states occurs via quantum jumps, the system cascades down the entropy ladder in steps, each releasing a finite amount of entropy. The Minkowski breaking at \(n=0\) marks the final step, where the inner horizon disappears and the classical geometry becomes discontinuous, signalling the impossibility of a smooth transition to the Schwarzschild point mass. Within the standard semiclassical framework, the transition probability across this point is exponentially enhanced by the positive entropy difference, in full compliance with the second law of thermodynamics.

These findings open several promising avenues for future investigation. First, a rigorous computation of the Euclidean action for the static regular black hole solutions would test the expected tunneling rate and confirm the thermodynamic preference for the Schwarzschild ground state.  Second, the analysis should be extended to the \(N>1\) families of the OCK construction, where the richer internal structure could lead to a more complex entropy spectrum and possibly to larger entropy releases. Third, the connection to area quantization deserves a deeper exploration: if the event horizon area is allowed to vary through Hawking radiation, the discrete steps in the inner horizon area might be accompanied by corresponding steps in the outer horizon area, leading to a natural realization of Bekenstein's area spectrum \(A \propto \ell_P^2\, N\).  Fourth, a complete treatment would embed the kinematic model into a full dynamical collapse scenario, solving the Einstein equations coupled to a realistic matter source, and following the quantum radiation emitted during the transition.  Such a study would reveal whether the entropy release appears as a burst of Hawking radiation or gravitational waves, offering potential observational signatures.

The transition from a regular black hole to a Schwarzschild black hole also involves the emergence of a curvature singularity, which alters the interior causal structure (the regular core is replaced by a singular point). Whether this alteration could be elevated to a genuine topology change in the sense of quantum gravity is an open question, but the suggestion that the thermodynamic quantum phase transition would imply a topological transition is intriguing and deserves further exploration.

As shown in the Appendix, the extension to the \(N=2\) family reveals a richer multi-stage structure, where both \(n_1\) and \(n_2\) transitions contribute to the entropy release. The inner horizon radius and entropy fraction depend on the full set of integer parameters, and the convergence to the \(N=1\) limit is algebraic (\(\propto 1/n_2\)) rather than exponential, leading to noticeable deviations even for moderately large \(n_2\sim 10\). This multi-stage cascade is a distinctive prediction of the \(N>1\) models and could produce a train of discrete pulses if the transitions occur via quantum jumps.

Finally, the divergence of the inner horizon surface gravity as \(n\to 0^{+}\) suggests that the final stages of inner horizon disappearance lie firmly in the quantum regime. As shown in Sec.~\ref{sec:cascade}, the semiclassical description breaks down when \(\kappa_{\rm inner} \sim 1/\ell_P\), which occurs at \(n \sim 1/\ln(h/\ell_P)\). For astrophysical black holes, this gives \(n \sim 0.01\), significantly larger than the naive estimate \(\ell_P/h \sim 10^{-38}\), because the divergence is logarithmic in the mass. At this stage, the back‑reaction of the emitted radiation cannot be neglected.  A natural framework to address this regime is the stochastic semiclassical Einstein--Langevin equation, which could capture the fluctuations around the Minkowski breaking point and their role in the information loss problem.  It is tempting to speculate that the discrete nature of the regular black hole states uncovered here may be a hint of a deeper quantum structure of spacetime, in which the gravitational Hilbert space is spanned by a countable family of geometric states, with black hole entropy emerging as the logarithm of the number of such states consistent with the macroscopic exterior.

\begin{acknowledgments}
	FSNL acknowledges support from the Funda\c{c}\~{a}o para a Ci\^{e}ncia e a Tecnologia (FCT) Scientific Employment Stimulus contract with reference CEECINST/00032/2018, and funding through the research grant UID/04434/2025.
	MER thanks Conselho Nacional de Desenvolvimento Cient\'ifico e Tecnol\'ogico - CNPq, Brazil, for partial financial support.
\end{acknowledgments}

\appendix

\section{The $N=2$ family: richer internal structure and multi‑stage entropy release}
\label{sec:N2}

While the $N=1$ solution already captures the essential physics of a regular black hole with a single quantised parameter, the full OCK construction allows for an arbitrary number of exponents $n_i$.  
The case $N=2$ is the first instance where the interplay between two independent integer parameters can be studied analytically and numerically, revealing how additional internal degrees of freedom affect the inner horizon, the entropy content, and the collapse cascade.  
Understanding $N=2$ is a necessary step towards the statistical interpretation of the full $N>1$ families, where the number of microstates grows combinatorially and the Bekenstein–Hawking entropy may be recovered from a count of distinct geometric configurations.

For $N=2$ the interior Misner–Sharp mass function~\eqref{m_N} reduces to
\begin{multline}
	m(r) = \frac{r}{2}\Bigg[
	\frac{(n_1+1)(n_2+1)}{(n_1-2)(n_2-2)}\left(\frac{r}{h}\right)^{\!2} \\
	+ \frac{3(n_2+1)}{(n_1-2)(n_1-n_2)}\left(\frac{r}{h}\right)^{\!n_1}
	+ \frac{3(n_1+1)}{(n_2-2)(n_2-n_1)}\left(\frac{r}{h}\right)^{\!n_2}
	\Bigg],
	\label{m_N2}
\end{multline}
with integers $2 < n_1 < n_2$.  The metric function inside the event horizon retains the form $f(r)=1-2m(r)/r$, i.e.
\begin{equation}
	f(x)=1-A x^{2}+B x^{n_1}-C x^{n_2},\qquad x\equiv\frac{r}{h},
	\label{f_N2}
\end{equation}
where the positive coefficients are
\begin{align}
	A &= \frac{(n_1+1)(n_2+1)}{(n_1-2)(n_2-2)},\quad
	B = \frac{3(n_2+1)}{(n_1-2)(n_2-n_1)},\nonumber\\
	C &= \frac{3(n_1+1)}{(n_2-2)(n_2-n_1)}.
	\label{coeffs}
\end{align}
The inner (Cauchy) horizon $h_{\rm c}=h_{\rm c}(n_1,n_2)$ is the smallest positive root of $f(x)=0$ besides the event horizon $x=1$, and the total entropy of the regular black hole is again
\begin{equation}
	S_{\rm reg}(n_1,n_2) = \frac{A}{4} + \pi h_{\rm c}^{\,2}(n_1,n_2).
	\label{Sreg_N2}
\end{equation}

\medskip
\noindent
\textit{Geometry of the interior and the de~Sitter core.}
Near the origin the mass function behaves as
\begin{equation}
	m(r) \simeq \frac12\frac{(n_1+1)(n_2+1)}{(n_1-2)(n_2-2)}\,\frac{r^{3}}{h^{2}},
\end{equation}
so the central core is again a regular de~Sitter region with effective cosmological constant
\begin{equation}
	\Lambda_{\rm eff}(n_1,n_2) = \frac{3}{h^{2}}\frac{(n_1+1)(n_2+1)}{(n_1-2)(n_2-2)}.
	\label{Lambda_N2}
\end{equation}
When both exponents are large, $n_1,n_2\gg1$, $\Lambda_{\rm eff}\simeq 3/h^{2}$ and the metric approaches the pure de~Sitter form $f(r)\simeq 1-r^{2}/h^{2}$, exactly as in the $N=1$ case.  
If only $n_2$ is large while $n_1$ remains moderate, the core curvature is enhanced relative to the $N=1$ solution with the same $n_1$, and the inner horizon is pushed inward.

\medskip
\noindent
\textit{Asymptotics of the inner horizon: slow approach to the $N=1$ limit.}
When $n_2\gg n_1$, the coefficients~\eqref{coeffs} become
\begin{align}
	A &= \frac{n_1+1}{n_1-2}\Bigl[1+\mathcal{O}(n_2^{-1})\Bigr],\qquad
	B = \frac{3}{n_1-2}\Bigl[1+\mathcal{O}(n_2^{-1})\Bigr],\nonumber\\
	C &= \frac{3(n_1+1)}{n_2^{2}}\Bigl[1+\mathcal{O}(n_2^{-1})\Bigr] = \mathcal{O}(n_2^{-2}).
\end{align}
Because the term $C x^{n_2}$ is exponentially suppressed for any $x<1$, it contributes negligibly to the inner horizon.  
The leading corrections to the metric function stem from the $1/n_2$ deviations of $A$ and $B$, so that
\begin{equation}
	f(x) = f_{N=1}(x;n_1) + \mathcal{O}(n_2^{-1}),
\end{equation}
where the corrections are power‑law suppressed, not exponential.  Consequently, the inner horizon radius satisfies
\begin{equation}
	h_{\rm c}(n_1,n_2) = h_{\rm c}^{(1)}(n_1)\Bigl[1 + \frac{a(n_1)}{n_2} + \mathcal{O}(n_2^{-2})\Bigr],
	\label{eq:large_n2_correct}
\end{equation}
with $a(n_1)$ a negative coefficient of order unity.  
The convergence to the $N=1$ limit is therefore algebraic and rather slow; for astrophysically relevant values $n_2\sim 10$ the inner horizon can still differ noticeably from the single‑exponent case, as the numerical examples below show.

\medskip
\noindent
\textit{Numerical inner horizons and entropy fractions.}
Table~\ref{tab:N2_entropy} lists the exact inner horizon radii and entropy fractions for several representative pairs $(n_1,n_2)$, obtained by solving $f(x)=0$ numerically.  
The last column gives the $N=1$ value for comparison.  
Notice that for modest $n_2$ the inner horizon is significantly smaller than the corresponding $N=1$ horizon, and the approach to the $N=1$ result is gradual as $n_2$ increases.

\begin{table}[t]
	\caption{Inner horizon radius $x_{\rm c}=h_{\rm c}/h$ and entropy fraction $\Delta S/(A/4)=x_{\rm c}^{2}$ for $N=2$ regular black holes.  The $N=1$ limit $h_{\rm c}^{(1)}/h$ is approached only as $n_2\to\infty$.}
	\label{tab:N2_entropy}
	\begin{ruledtabular}
		\begin{tabular}{c c c c c}
			$n_1$ & $n_2$ & $h_{\rm c}/h$ & $\Delta S/(A/4)$ & $h_{\rm c}^{(1)}/h$ \\
			\hline
			3 & 4   & 0.5000 & 0.2500 & 0.7676 \\
			3 & 5   & 0.5353 & 0.2865 & 0.7676 \\
			3 & 7   & 0.585  & 0.342  & 0.7676 \\
			3 & 10  & 0.634  & 0.402  & 0.7676 \\
			3 & 100 & 0.762  & 0.581  & 0.7676 \\
			\hline
			4 & 5   & 0.586  & 0.343  & 0.8165 \\
			4 & 7   & 0.630  & 0.397  & 0.8165 \\
			4 & 10  & 0.675  & 0.456  & 0.8165 \\
			4 & 100 & 0.813  & 0.661  & 0.8165 \\
			\hline
			5 & 7   & 0.665  & 0.442  & 0.8484 \\
			5 & 10  & 0.707  & 0.500  & 0.8484 \\
			5 & 100 & 0.844  & 0.712  & 0.8484 \\
		\end{tabular}
	\end{ruledtabular}
\end{table}

The table illustrates two important features.  
First, the inner horizon area—and therefore the hidden entropy—can be substantially smaller than in the $N=1$ case when $n_2$ is not extremely large.  
Second, the sensitivity to $n_2$ decreases as $n_1$ increases: for $n_1=5$ the horizon is already at $0.707$ for $n_2=10$, while for $n_1=3$ it is only $0.634$.  
This reflects the fact that a larger $n_1$ itself already brings the geometry close to the de~Sitter limit, leaving less room for $n_2$ to influence the core.

\medskip
\noindent
\textit{Multi‑stage entropy cascade.}
The existence of two independent integer parameters dramatically enriches the collapse dynamics.  
During gravitational collapse the null convergence condition forces both exponents to decrease, preserving the ordering $n_1(v)<n_2(v)$~\cite{Ovalle:2026lxb}.  
The smaller exponent $n_1$ is the first to reach the threshold $n_1=2$ where a logarithmic curvature singularity appears; the inner horizon, however, survives until $n_1=0$, when Minkowski breaking occurs and the inner horizon vanishes.  
Therefore the entire entropy release $\Delta S = A/4$ must be accomplished while $n_1>0$.

Within this regular phase, the entropy release depends on the nature of the evolution. During a classical continuous collapse, the parameters $n_i(v)$ evolve through all real values, so the entropy $S_{\rm reg}(n_1,n_2)$ and hence the entropy release are continuous functions of time. However, the static equilibrium states are labelled by integer values of $(n_1,n_2)$; if the transition between these equilibrium states occurs via quantum jumps, the entropy would be released in a sequence of finite drops as the discrete parameters pass through integer values. Both $n_1$ and $n_2$ transitions are significant:
\begin{enumerate}
	\item \textbf{$n_1$‑transitions:} $n_1\to n_1-1$ with $n_2$ fixed.  
	The entropy drop can be read directly from Table~\ref{tab:N2_entropy}.  
	For example, $(5,10)\to(4,10)$ releases $\approx 0.044\,(A/4)$, and $(4,10)\to(3,10)$ releases $\approx 0.054\,(A/4)$.
	\item \textbf{$n_2$‑transitions:} $n_2\to n_2-1$ while $n_1$ is kept fixed.  
	These steps can be equally substantial.  
	As $n_2$ decreases from $10$ to $4$ with $n_1=3$, the entropy fraction drops from $0.402$ to $0.250$, a release of $0.152\,(A/4)$ over several steps.
\end{enumerate}

In principle, simultaneous transitions in which both $n_1$ and $n_2$ cross integer values at the same instant are also possible, provided the ordering constraint $\dot n_1(v) \le \dot n_2(v)$ is satisfied. Such a simultaneous transition would release the sum of the individual entropy drops in a single pulse. For example, a transition $(n_1,n_2)\to(n_1-1,n_2-1)$ would release an amount $\delta S = [S_{\rm reg}(n_1,n_2)-S_{\rm reg}(n_1-1,n_2)] + [S_{\rm reg}(n_1-1,n_2)-S_{\rm reg}(n_1-1,n_2-1)]$, evaluated at the transition point. For simplicity, we consider the sequential picture here, which captures the essential multi-stage structure and is sufficient to illustrate the qualitative features of the entropy cascade.

A complete cascade starting from a highly excited state with $n_1,n_2\gg1$ (where $S_{\rm reg}\approx 2A/4$) will therefore consist of a series of $n_1$ and $n_2$ transitions, each releasing a fraction of $A/4$, until the inner horizon finally vanishes at $n_1=0$.  
The total liberated entropy is always $A/4$, independent of the path.  
This multi‑stage structure is a distinctive signature of the $N>1$ models: if the entropy release manifests as radiation, one expects a train of discrete pulses whose amplitudes are determined by the integer jumps.  
Such a pattern could serve as an observational discriminant between different regular black hole scenarios.

\medskip
\noindent
\textit{Statistical mechanics of the interior.}
The presence of $N$ ordered integers naturally discretises the internal geometry.  
For fixed ADM mass $\mathcal{M}$, the number of regular microstates $\Omega(\mathcal{M})$ is the number of $N$‑tuples $(n_1,\dots,n_N)$ with $2<n_1<\dots<n_N\le n_{\rm max}$, where $n_{\rm max}$ is a cutoff imposed by quantum gravity (see Sec.~\ref{sec:quantization}).  
For $N=1$ this number grows linearly with $n_{\rm max}$; for $N=2$ it becomes the binomial coefficient $\binom{n_{\rm max}-2}{2}\approx n_{\rm max}^{2}/2$, drastically increasing the state space for the same macroscopic mass.  
In the limit $N\to\infty$ and $n_{\rm max}\gg1$, the number of microstates grows exponentially, and the Bekenstein–Hawking entropy $S_{\rm BH}=A/4$ emerges as the logarithm of $\Omega(\mathcal{M})$ provided $n_{\rm max}$ scales appropriately with $\mathcal{M}$.  
This offers a concrete realisation of the proposal that black hole entropy counts the internal geometric degrees of freedom that are indistinguishable from the outside: each regular black hole with a given set $\{n_i\}$ is a distinct microstate, and the thermodynamic entropy is the logarithm of the number of such states compatible with the exterior Schwarzschild geometry.




\begin{thebibliography}{99}
	
	\bibitem{Penrose:1964wq}
	R.~Penrose,
	``Gravitational collapse and space-time singularities,''
	Phys. Rev. Lett. \textbf{14} (1965), 57-59.
	
	\bibitem{Hawking:1970zqf}
	S.~W.~Hawking and R.~Penrose,
	``The Singularities of gravitational collapse and cosmology,''
	Proc. Roy. Soc. Lond. A \textbf{314} (1970), 529-548.
	
	\bibitem{Hawking:1973uf}
	S.~W.~Hawking and G.~F.~R.~Ellis,
	``The Large Scale Structure of Space-Time,''
	Cambridge University Press, Cambridge, 1973.
	
	\bibitem{Bardeen:1968qtr}
	J.~M.~Bardeen,
	``Non-singular general-relativistic gravitational collapse,''
	in \textit{Proceedings of the International Conference GR5}, Tbilisi, USSR, 1968, p.~174.
	
	\bibitem{Dymnikova:1992ux}
	I.~Dymnikova,
	``Vacuum nonsingular black hole,''
	Gen. Rel. Grav. \textbf{24} (1992), 235-242.
	
	\bibitem{Hayward:2005gi}
	S.~A.~Hayward,
	``Formation and evaporation of regular black holes,''
	Phys. Rev. Lett. \textbf{96} (2006), 031103
	[arXiv:gr-qc/0506126 [gr-qc]].
	
	\bibitem{BenAchour:2020gon}
	J.~Ben Achour, S.~Brahma, S.~Mukohyama and J.~P.~Uzan,
	``Towards consistent black-to-white hole bounces from matter collapse,''
	JCAP \textbf{09} (2020), 020
	[arXiv:2004.12977 [gr-qc]].
	
	\bibitem{Ovalle:2024wtv}
	J.~Ovalle,
	``Schwarzschild black hole revisited: Before the complete collapse,''
	Phys. Rev. D \textbf{109} (2024) no.10, 104032
	[arXiv:2405.06731 [gr-qc]].
	
	\bibitem{Ovalle:2026lxb}
	J.~Ovalle, R.~Casadio and A.~Kamenshchik,
	``Schwarzschild black hole singularity formation,''
	Phys. Rev. D \textbf{113} (2026) no.6, 064042
	[arXiv:2603.06451 [gr-qc]].
	
	\bibitem{Wald:1993nt}
	R.~M.~Wald,
	``Black hole entropy is the Noether charge,''
	Phys. Rev. D \textbf{48}, no.8, R3427-R3431 (1993)
	[arXiv:gr-qc/9307038 [gr-qc]].
	
	\bibitem{Iyer:1994ys}
	V.~Iyer and R.~M.~Wald,
	``Some properties of Noether charge and a proposal for dynamical black hole entropy,''
	Phys. Rev. D \textbf{50}, 846-864 (1994)
	[arXiv:gr-qc/9403028 [gr-qc]].
	
	\bibitem{Wald:1999vt}
	R.~M.~Wald,
	``The thermodynamics of black holes,''
	Living Rev. Rel. \textbf{4}, 6 (2001)
	[arXiv:gr-qc/9912119 [gr-qc]].
	
	
	\bibitem{Hawking:1975vcx}
	S.~W.~Hawking,
	``Particle Creation by Black Holes,''
	Commun. Math. Phys. \textbf{43} (1975), 199-220
	[erratum: Commun. Math. Phys. \textbf{46} (1976), 206].
	
	\bibitem{Bekenstein:1974jk}
	J.~D.~Bekenstein,
	``The quantum mass spectrum of the Kerr black hole,''
	Lett. Nuovo Cim. \textbf{11} (1974), 467.




\end{thebibliography}
\end{document}